# Unusual giant magnetoresistance effect in heterojunction structure of ultra-thin single-crystal Pb film on silicon substrate

Jian Wang [a]

Institute of Physics, Chinese Academy of Sciences, Beijing 100080, China and The Center for Nanoscale Science and Department of Physics, The Pennsylvania State University, University Park, Pennsylvania 16802-6300, USA

Xu-Cun Ma, Yun Qi, Ying-Shuang Fu, Shuai-hua Ji, and Li Lu

Institute of Physics, Chinese Academy of Sciences, Beijing 100080, China

X. C. Xie

Institute of Physics, Chinese Academy of Sciences, Beijing 100080, China and Department of Physics, Oklahoma State University, Stillwater, OK 74078, USA

Jin-Feng Jia, Xi Chen, and Qi-Kun Xue [b]

Department of Physics, Tsinghua University, Beijing 100084, China

Superconductor films on semiconductor substrates draw much attention recently since the derived superconductor-based electronics have been shown promising for future data process and storage technologies. By growing atomically uniform single-crystal epitaxial Pb films of several nanometers thick on Si wafers to form a sharp superconductor-semiconductor heterojunction, we have obtained an unusual giant magnetoresistance effect when the Pb film is superconducting. In addition to the great fundamental interest of this effect, the simple structure and compatibility and scalability with current Si-based semiconductor technology offer a great opportunity for integrating superconducting circuits and detectors in a single chip.

[a] Electronic mail: juw17@psu.edu
[b] Electronic mail: qkxue@mail.tsinghua.edu.cn

The use of dissipationless superconducting components will produce denser and more rapid chips since the resistance of interconnecting metal circuits is a major source of heat generation and charging time.[1] Motivated by rapid progress in superconducting electronics,[2-5] such as logic circuits, sensitive detectors and nonvolatile memories, superconductor-semiconductor hybrid structures have become an attractive field in recent years.[6-10] In the late 1980's,[11, 12] two groups discovered the giant magnetoresistance (GMR) effect in the metal films composed of alternative ferromagnetic and nonmagnetic layers, in which the magnetizations of adjacent ferromagnetic layers can be switched between antiparallel and parallel state by external magnetic field, causing a substantial change in resistance. The discovery of GMR not only significantly improves our knowledge on spin-dependent electronic transport processes, but also leads to tremendous applications in the read heads in modern hard drivers and in non-volatile magnetic random access memories.[13, 14] A holy grail would be an integration of the GMR effect with Si-based microelectronic technology.

Here, we report our experimental observation of an unusual GMR phenomenon in simple Pb-Si superconductor-semiconductor heterojunctions. By growing ultra-thin single-crystal Pb films (<10nm in thickness) on Si(111) substrates, a giant negative magnetoresistance effect was observed when the Pb thin films become superconducting at low temperature, which is totally different with the large positive magnetoresistance in nonmagnetic materials.[15-18] Since the structure does not contain any magnetic element, the underlying mechanism is fundamentally different from that for the traditional GMR in metal multilayers.[12, 13]

Our Pb thin films were prepared on heavily doped $n^{++}$ Si(111) substrates by standard molecular beam epitaxy (MBE) technique.[19] During growth, the Si substrates were cooled down to 95 K by liquid nitrogen (LN$_2$) to achieve atomically smooth single-crystal Pb thin films, as reported elsewhere.[20-26] Figure 1(a) shows a typical scanning tunneling microscopy (STM) topographic image of the Pb thin film with a thickness of 26 atomic monolayers (ML), from which the atomically smooth nature of the film is immediately evident. Before the samples were taken out from the ultra-high vacuum growth chamber for transport property measurement, 4 ML Au was deposited on the film to protect contamination and surface oxidation in ambient condition.[20, 24] The transport measurements were carried out by using standard four electrodes method in a Physical Property Measurement System (Quantum Design-Model 6000). As shown in Fig. 1(b), the film exhibits a superconducting transition at 6.4 K (T$_C$), and no residual resistance was found.

To measure the transport properties through the Pb-Si heterojunctions, the film was cut into two parts with a 2 μm wide gap made by focused ion beam (Focused Ion Beam Etching & Depositing System, FEI-DB235) (see Fig. 1(c)). The etching current with Ga ions was less than 10 pA, contamination and damage of the structure by the Ga ions could be mostly avoided. Four indium electrodes with Au wires of 25 micrometers in diameter were made and connected to two parts of the film. The measurement geometry is schematically shown in the insert of Fig. 1(d). Because the resistances of both doped Si substrate and Pb film are very small (the resistance of the n$^{++}$ Si wafer used is below 0.1 Ω even at 2.5 K), the measurement mainly reflects the transport property of the two Pb-Si heterojunctions. Figure 1(d) shows the resistance-temperature (R-T) curve of this double-junction structure. Below 7.0 K the resistance drops slightly at first. Then, with further decreased temperature, the resistance increases rapidly.

Figures 2(a) and 2(b) show the measured resistance (R) as a function of the magnetic field (H) applied perpendicularly to the Pb film at different temperatures. It is clear that the resistance decreases rapidly (at an averaged rate of ~0.42 Ω/Oe for T=2.5 K) with increasing magnetic field, eventually reaching a plateau at the critical field $H_C$. It is also clear that both the maximal resistance and $H_C$ increase with decreasing temperature. At T=5.5 K, the resistance decreases by a factor of 1.3 when H is increased from 0 to 0.9 kOe. Remarkably, that factor increases to 3.1 with a field change of 2.6 kOe at T=2.5 K, in comparison with a factor of about 2 for the traditional GMR effect in the Fe/Cr system under a field of 20 kOe at 4.2 K[11]. Besides the large peak, the resistance also exhibits a weak minimum at a magnetic field just below $H_C$. The resistance minimum becomes more pronounced with increasing temperature (below 7 K). The phenomenon was verified on several samples. Unlike the GMR effect in the metallic multilayers caused by the interlayer magnetic coupling,[11, 12] the GMR effect observed here will vanish above 7 K.

For comparison, R-H scan of the 26 ML Pb film is shown in Figure 2(c). From this figure one can clearly see that, at the same temperature the upper critical field $H_{C2}$ of the Pb film is much larger than the $H_C$ of the Pb-Si junctions (see Figs 2(a), 2(b)). For example, at 2.5 K, the $H_{C2}$ of the Pb film is 7 kOe, while the corresponding $H_C$ of the Pb-Si junctions is only 2.8 kOe. It is surprising that between 7 kOe and 2.8 kOe, although the Pb film is still superconducting, the GMR effect no longer exists.

Figure 3(a) shows more details of the R-H curves of the Pb-Si junctions. Sharp valley-like resistance minima are found from 2.5 K to 6.8 K. With increasing temperature, the large resistance peak at zero field gradually fades away and the valley-like resistance minima are approaching to the zero field. At 7.0 K the peak basically disappears and the resistance minimum is at zero field. The R-H curve becomes a straight line above 7.0 K. The data clearly reveal that the resistance of the Pb-Si junctions is very sensitive to the temperature when temperature is below 7.0 K. A very small change (0.1 K) in temperature could induce more than 10 Ω change in resistance. Such sensitivity makes the Pb-Si junction highly promising for mass-production of new cryogenic temperature-detectors on Si chips.[4, 27]

In order to further understand the effect of the superconducting Pb film on the observed GMR effect, differential conductance experiments were carried out with the same sample. Figure 3(b) shows the dI/dV-V curves measured under various magnetic fields at 2.5 K. The superconducting gap owing to BCS-like density of states[28] is visible at zero magnetic field, and gradually disappears as the magnetic field is raised. Since the result is from two Pb-Si junctions, the width between the two BCS-like peaks is twice that of the superconducting gap. Hence, the gap is 3.3 meV, which is larger than the superconducting gap of bulk Pb, for which the gap is 2.73 meV at 0 K. In Figure 3(c), we plot the differential resistance versus the magnetic field at zero voltage, which is obtained from the data of dI/dV-V in Figure 3(b). We can see the almost same GMR behavior as observed in R-H measurements (Figs 2(a), 2(b)).

Recently, negative magnetoresistance in disordered thin films and wires has also been observed.[29-32] However, the enhanced negative magnetoresistance (GMR) behavior found in heterojunctions of single-crystal superconductor film and semiconductor substrate has never been reported before. One possible source of the unusual GMR behavior is the electronic tunneling in a superconductor-normal metal (S-N) junction,[33] namely, the tunneling between the superconducting Pb film and n$^{++}$ Si substrate through the Schottky-barrier[34] at the epitaxial

Pb/Si(111) interface. With an increasing magnetic field, the electron density for tunneling increases as shown in Fig. 3(b), thus, the S-N-like tunneling is enhanced with increasing field. Accordingly, the resistance of the junction decreases with an increasing field. Nevertheless, this simple picture does not explain the finding that the $H_C$ of the structure is much less than the upper critical field $H_{C2}$ of the Pb film. It is also of difficulty to understand the fact that the resistance exhibits a weak minimum at H just below $H_C$.

Another possible qualitative explanation is from BTK model[35] and Andreev reflection[36]. We know that there is a tunneling barrier between the Pb film and the Si substrate due to formation of the Schottky-barrier at the interface. Because the resistance of the sample is not too big (above 2 K, the resistance is below $2k\Omega$, see Fig. 1(d)), we believe that the strength of the barrier is intermediate (between zero barrier and a strong tunnel barrier). Blonder *et al.*[35] (BTK) introduced a $\delta$-function potential barrier of strength Z at the interface to study the electric tunneling. If the strength Z is zero, there is no barrier at the interface between the normal and the superconducting metals. The electrical current transfer process is a novel reflection process described by Andreev.[36] This situation applies to a normal metal-superconductor junction. If the Z is very large (for example lager than 10), there is a classic high barrier tunnel junction and electron tunneling dominates the electron transport. For our sample, the Z is not zero, but is not large either. According to the BTK model, for this situation, the probability of Andreev reflection is increased when the electron energy is changing from 0 to $\Delta(T)$ ($2\Delta$ is the superconducting gap according to BCS theory[28]), then, it decreases rapidly with a further increased electron energy, as sketched in the insert in Fig. 3(c). Since the effective superconducting gap decays with increase of an applied magnetic field, the electron energy in our measurement becomes close to $\Delta$ with increasing field. Accordingly, the probability of Andreev reflection increases and the resistance decreases. When the electron energy equals $\Delta$, the probability of Andreev reflection reaches a maximum value, correspondingly, the resistance reaches a minimum in the R-H curves. However, the BTK model and Andreev reflection can not explain the fact that we got almost same GMR behavior by using 50 nA and 500 nA currents for the measurement, since the energy of the electron in the 500 nA measurement current is 0.8 meV at 2.5 K and zero field and the superconducting gap $2\Delta$ (2.5 K) is 3.3 meV.

We do not as of yet have a satisfactory model to explain the unusual GMR effect found in Pb-Si structure. Maybe the formation of quantum well states, which greatly modulates the electronic structure near the Fermi energy[20-22, 24, 37, 38] in the present Pb film, also plays an important role. We expect that our work will stimulate further theoretical studies.

The unusual GMR effect in superconductor-semiconductor heterojunction may be utilized for developing a magnetic-field controlled "on-off" device or a high-sensitivity field sensor. Because it is from the electron transport across the Pb/Si(111) interface, fabrication of any devices based on the effect could be scaled up for mass manufacture using the well-established microelectronics technology. This effect may also be utilized, or need to be avoided in some cases, in the future hybrid circuits of the traditional microelectronics and the emerging superconducting quantum-electronics.[39]

This work was financially supported by National Science Foundation and Ministry of Science and Technology of China.

**Figure legends**

**Figure 1.** (a) A scanning tunneling microscope image of the 26 ML atomically flat Pb thin film. (b) R vs T curve measured from the Pb film shown in Fig. 1(a), showing that a superconductivity transition at a temperature of 6.4 K. (c) A scanning electron micrograph of the Pb film after a 2 μm wide gap (the dark region) was fabricated. (d) R vs T obtained from the Pb-Si-Pb double-junction structure. The inset is the schematic graph for the transport measurement across the Pb/Si(111) heterojunctions.

**Figure 2.** (a) Magnetoresistance of the heterojunctions with a magnetic field perpendicular to the film at different temperatures. (b) Close-view of Fig. 2(a) near zero magnetic field for clarity. The vertical scale is normalized to the resistance at zero magnetic field. Note that there is no GMR effect when the film is in the normal state (the pink line). (c) R vs H curves of the 26 ML Pb film with a magnetic field perpendicular to the film at indicated temperatures.

**Figure 3.** (a) R-H curves of the heterojunctons with a magnetic field perpendicular to the film at 6.5 K, 6.6 K, 6.7 K, 6.8 K, 6.9 K, and 7.0 K, respectively. (b) Differential conductance dI/dV vs voltage V curves of the heterojunctons at indicated magnetic fields at 2.5 K. (c) dV/dI vs H curve of the heterojunctons at 2.5 K when the applied voltage is zero. The data are from dI/dV-V curves. The inset is from the paper of Blonder *et al*.[30] It shows differential conductance for barrier strength Z=0.5 at T=0 K.

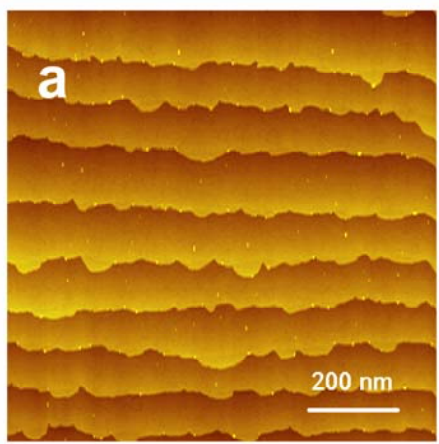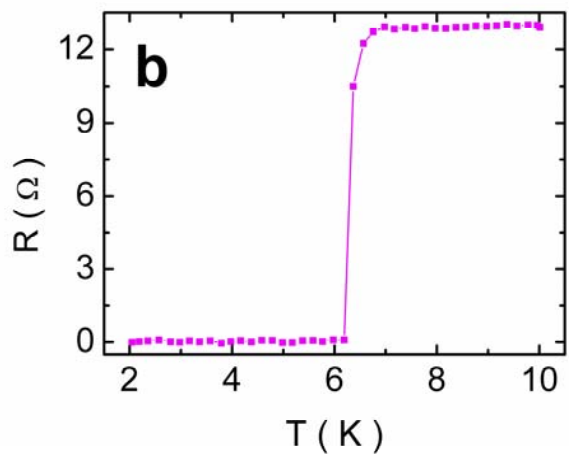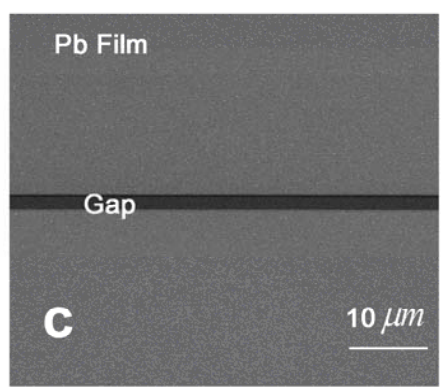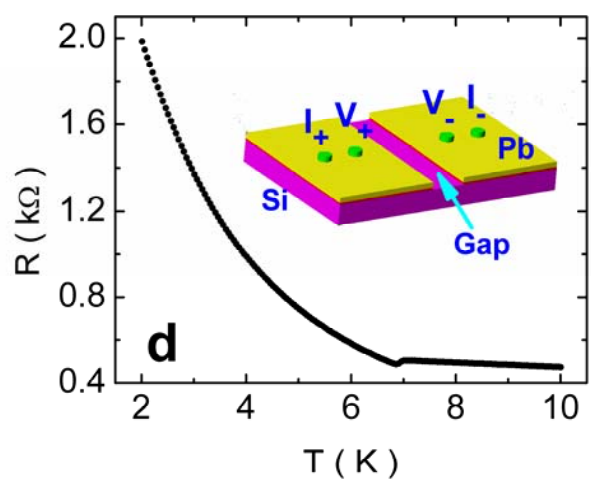

Wang et al., FIG. 1

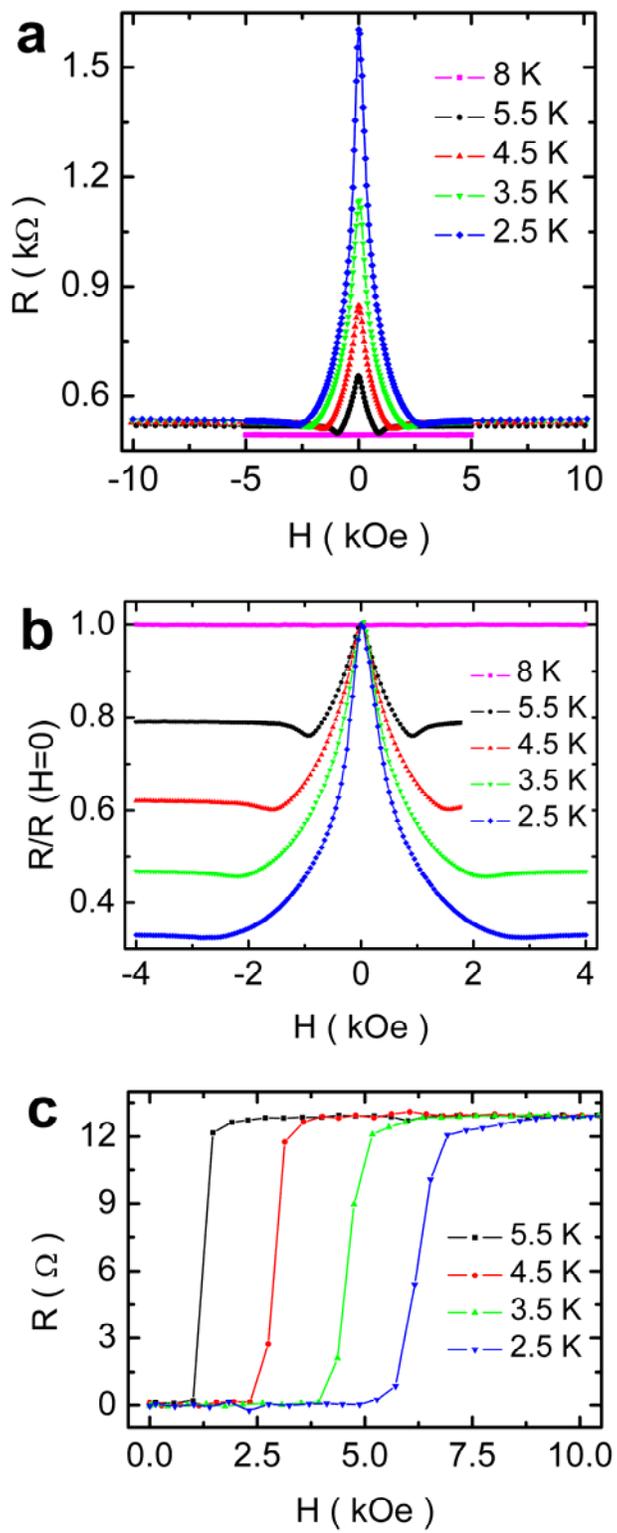

Wang et al., FIG. 2

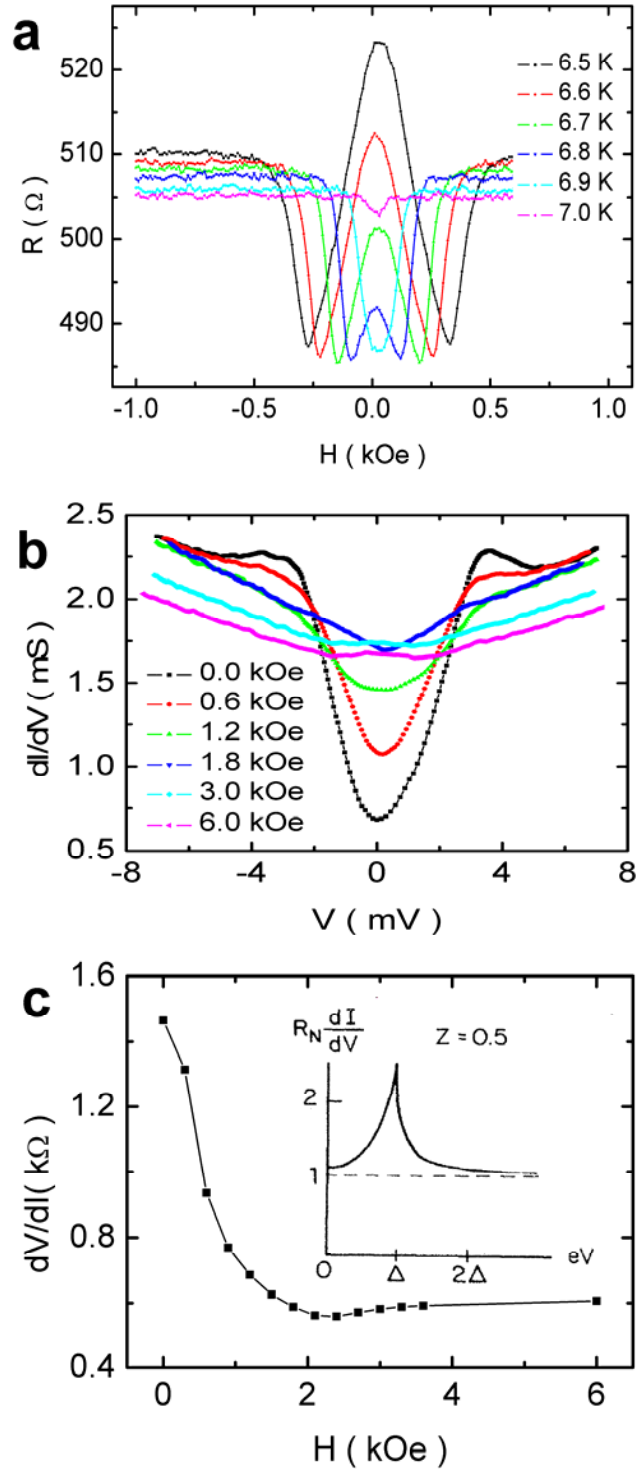

Wang et al., FIG. 3